\pdfoutput=1

\documentclass[twoside]{article}
\usepackage{graphicx}
\usepackage{titling}
\usepackage{setspace}
\usepackage{authblk} 
\usepackage{graphicx}
\usepackage{amsbsy} 
\usepackage{amsmath} 
\usepackage{siunitx} 
\usepackage{amstext} 
\usepackage[hang,flushmargin]{footmisc} 
\usepackage{float} 
\usepackage{gensymb} 
\usepackage{url}
\usepackage{amssymb} 
\usepackage{caption} 
\usepackage{subcaption} 

\usepackage[font={small}]{caption} 
\usepackage[margin=1in]{geometry}

\title{WAVELENGTH AND REFRACTIVE INDICES FROM INTERFEROMETERY}
\author{Kam Modjtahedzadeh}
\affil{Department of Physics, University of California, Santa Barbara}
\date{August 29, 2019}

\begin{document}
\clearpage\maketitle
\thispagestyle{empty}

\vspace{0.8cm}

\begin{changemargin}{0.6in}{0.6in} 
\section*{Abstract} 
PASCO scientific 012-05187C Precision Interferometer is used in Michelson mode to investigate wavelengths and refractive indices.
From varying the distance of the movable mirror in the Michelson setup the wavelength of the HeNe laser beam is found to be $630.6\pm7.9~\mathrm{nm}$; $0.28$ sigmas away from the accepted 632.8 nm and agreeing with it. Then after considering the fact that the index of refraction for low pressure gasses varies linearly with pressure we place a vacuum cell in front of the movable mirror and pump out the air within it to find the individual slopes. By extrapolating the average slope we calculate the index of refraction for air to be $n=1.000226\pm0.000026$. This is 1.44 sigmas away from the manufacturer's measured 1.000263 and barely agreeing with it as we underestimated our error in the fringe count (which is caused by the change in pressure). Furthermore, the vacuum cell is replaced by a crown glass plate which is rotated to vary the length at which the EM wave travels in the Michelson interferometer. The angle of rotation is measured and utilized to find the refractive index of glass to be $1.514\pm0.006$; agreeing with the accepted value of 1.515 as it is 0.125 sigmas away from it. Thus, along with the wavelength observation this experiment is conducted successfully. Although the experiment to find the refractive index of air is conducted less successfully, it is also prosperous as the measured value is close to the accepted.
\end{changemargin}

\newpage
\twocolumn
\section{Introduction}

Interferometers are laser based tools used to investigate various properties directly or indirectly relating to EM waves. In our experiment, we resort to the widely used Michelson mode setup to measure the wavelength of our laser beam as well as the indices of refraction for air and crown glass.
\vspace{0.05cm} 
\begin{figure}[H]
\centering
\includegraphics[scale=0.3]{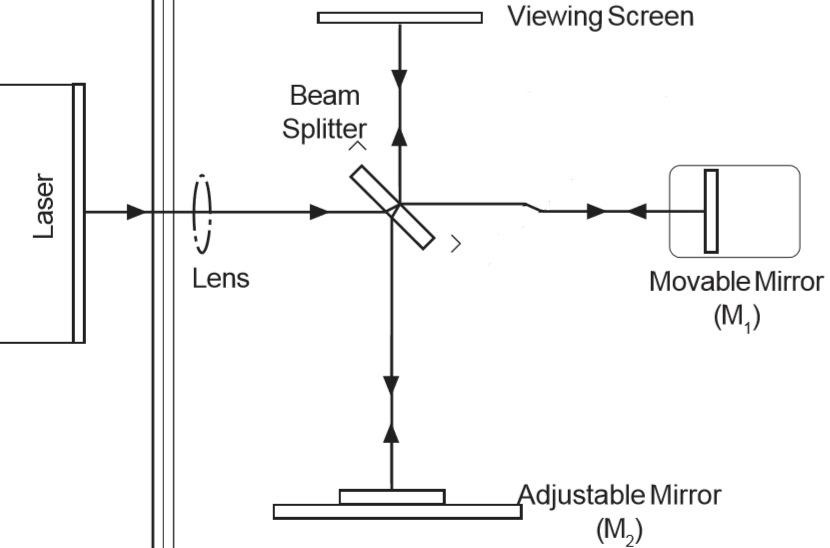}
\caption{A schematic of the Michelson interferometer. As the beam exits the laser source it passes through an $18~\mathrm{mm}$ lens and hits the beam splitter. Part of the beam goes through and reflects back after hitting a movable mirror $\mathrm{M}_1$; the other part reflects to an adjustable mirror $\mathrm{M}_2$, which also reflects it back to the beam splitter. Both parts of the split beam make their way to the viewing screen and create visible fringes.}
\label{fig:1}
\end{figure} \par
To find the wavelength of our laser beam, we will be utilizing Bragg's Law:
\begin{align}
2d_m\sin\Theta=N\lambda\,,
\label{1}
\end{align}
where $\lambda$ is the wavelength, $d_m$ is the distance that $\mathrm{M}_1$ moves toward the beam-splitter,\footnote{Hence the subscript $m$.} and $N$ is the number of the counted fringes.
\begin{figure}[H]
\centering
\includegraphics[scale=0.3]{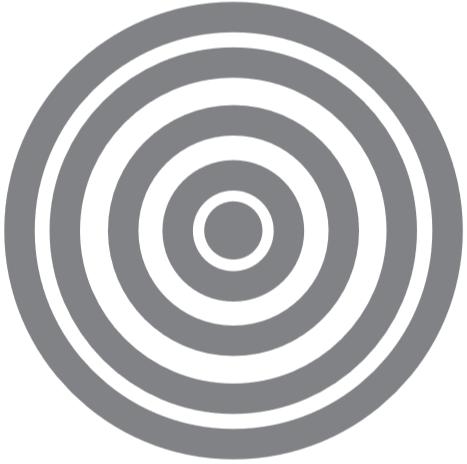}
\caption{Fringe pattern as seen on viewing screen.}
\label{fig:2}
\end{figure}
\noindent In our case $\Theta$ is 90\degree (as seen in Figure \ref{fig:1}); therefore,
\begin{align}
\lambda=\dfrac{2d_m}{N}\,.
\label{2}
\end{align}
So after measuring $d_m$ and $N$ we can calculated $\lambda$.\par

After measuring $\lambda$ we move on to find the refractive indices of air and glass, and start with that of air, $n_a$. To understand how to measure this, it is crucial to know that for EM waves:
\begin{align}
    \lambda=\dfrac{\lambda_0}{n}\,,
    \label{3}
\end{align}
where $\lambda_0$ is the wavelength of the EM wave in vacuum and $n$ is the index of refraction of the media it is traveling through \cite{Wavelength}. Knowing \eqref{3}, we attempt to find $n_a$ by constructing an alternative media in the midst of our Michelson interferometer. We do this by placing a vacuum cell between the beam-splitter and M$_1$. It is also critical to know that for low pressure gasses the index of refraction varies linearly with pressure \cite{nvsP}. Combining this knowledge with \eqref{2} we can calculate $n_a$ after measuring the manual change in cell pressure $\Delta P$ and observing its affect on the fringe pattern.\par

Furthermore, since glass is rigid we must use an alternative method to measure its index of refraction $n_g$. So we use the method of varying the length of the crown glass\footnote{We place a glass plate in the same location that the vacuum cell was previously placed. More on this in section 2.3.} which our laser passes through by revolving it using a rotating table and movable arm. The change in path length and measured $N$ relate to $n_g$ by the following equation:
\begin{align}
    N=\dfrac{2n_ad_a(\theta)+2n_gd_g(\theta)}{\lambda_0}\,,
    \label{4}
\end{align}
where $d_a(\theta)$ and $d_g(\theta)$ are the changes of path length in air and glass respectively \cite{3}. Since they are functions of the angle $\theta$ which the glass rotates, we can find $n_g$ if we measure $\theta$ and $N$.

\section{Methods}
To conduct our experiments we used the PASCO scientific 012-05187C Precision Interferometer which included a non-polarized yet standard HeNe laser with a wavelength of $\lambda_0=623.8$ nm \cite{INTERFEROMETRY}.\footnote{We re-measure the laser's wavelength, of course.}

\subsection{Distance Measurement}
The procedure of measuring the beam wavelength starts after we  set our interferometer in the Michelson mode and align the laser. Behind M$_2$ there is a micrometer which actually moves $\mathrm{M}_1$ back and forth; by setting it to 50 microns M$_1$ and the micrometer reading have a virtually linear relationship. We then adjust the viewing screen such that our chosen reference fringe is centered on it. Now by increasing the micrometer reading we see a shift in the fringe pattern; so we record the number of fringes which pass our reference mark as well as the distance moved by the micrometer, which is, $$d_m = \left|\mathrm{final~position - initial ~position}\right|\,.$$ A figure of our measurements is plotted below.
\vspace{-0.3cm}
\begin{figure}[H]
\centering
\includegraphics[scale=0.5]{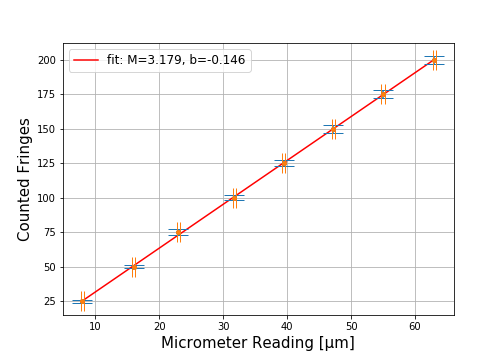}
\caption{Measured $N~\text{vs}~d_m$. Vertical error bars correspond to the uncertainty in $N$, horizontal bars correspond to the uncertainty in $d_m$. Using the SciPy package in Python, a fitted line $M\cdot d_m+b$ is included. The variance of the parameters are found by taking the square root of the SciPy variance function \textit{pcov[]}; $\delta M=0.019$ and $\delta b=0.755$. If the relationship between M$_1$ and the $d_m$ was not linear, the fit might not have been a regression line.}
\label{fig:3}
\end{figure}
\noindent The uncertainty in all micrometer measurements $\delta d_m$ is $0.25$ microns, as the dial spacings are very tiny. The uncertainty for $N$ varies by the number of fringes we counted, since the more you count the easier it can be to lose track of the rings in Figure \ref{fig:2}. 
\vspace{0.5cm}
\begin{table}[H]
\centering
\caption{Uncertainties for $N$ Depending on the Observed Rings}
\label{tab:1}
\begin{tabular}{c|c}
$N$ & $\delta N$\\ \hline
\multicolumn{1}{c|}{$1\leq N\leq50$} & 1 \\
\multicolumn{1}{c|}{$50< N<150$} & 2 \\
\multicolumn{1}{c|}{$150\leq N\leq200$} & 3 \\
\end{tabular}
\end{table}

\subsection{Vacuum Cell Pressure Change}
We now place a vacuum cell before M$_1$ as explained in the introduction. We measure the width of our cell to be $d=2.95\pm0.01$ cm.\footnote{According to the manufacturer PASCO it should be $3$ cm, this is incorrect!} The uncertainty comes from the fact that it is difficult to measure centimeters up to two decimal places with full accuracy.
\begin{figure}[H]
\centering
\includegraphics[scale=0.3]{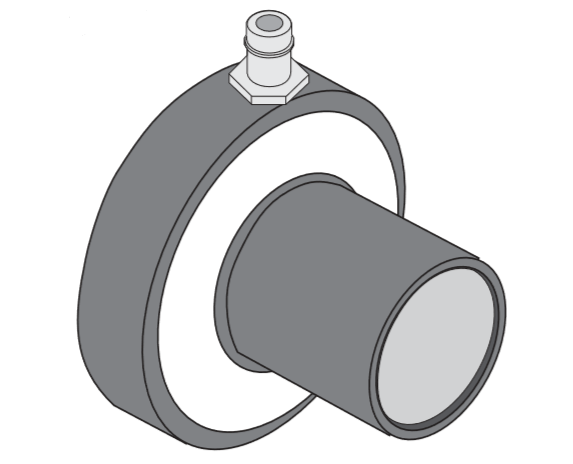}
\caption{Image of our vacuum cell. The part number is 003-05162 from PASCO. The thickness $d$ is measured from both the part with the larger area and the one with the lower.}
\label{fig:4}
\end{figure}
\noindent It is important to know $d$ because changing the travel length of our EM wave in an alternative media has the same effect as $d_m$ in \eqref{2}. Moreover, after decreasing the pressure in the cell with a vacuum pump\footnote{The OS-8502 Hand-Held Vacuum Pump, to be exact.} we record the change in pressure\footnote{The reading on the pump is $\Delta P$.} $\Delta P$ as well as the number of fringes that pass through our reference point. 
\begin{figure}[H]
\centering
\includegraphics[scale=0.5]{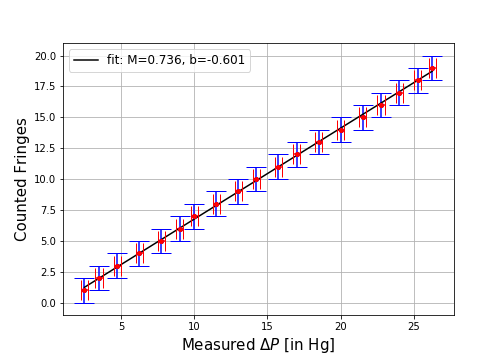}
\caption{Measured $N~\text{vs}~\Delta P$, counted fringes increase linearly with pressure. As displayed, pressure was measured in units of in Hg; but, it was converted to Pascals for analysis. Based on the measured units a curve $M\cdot\Delta P+b$ is fitted with the parameters given in the legend. The  square root of their variances are $\delta M=0.004$ and $\delta b = 0.071$.}
\label{fig:5}
\end{figure}
\noindent The uncertainty in $\Delta P$ comes from the vacuum pump reading and is a constant $0.25~\mathrm{in~Hg}$. There were no fluctuations in the reading as the device was calibrated shortly before we used it; hence the small error. See Table \ref{tab:1} for the error in $N$.

\vspace{-1.5cm}\subsection{Angle Quantification}
Replacing the vacuum cell with a crown glass plate sitting on a rotating table, we have the tools necessary to get the measurements for the refractive index of glass.
\begin{figure}[!htbp]
  \centering
  \includegraphics[scale=0.3]{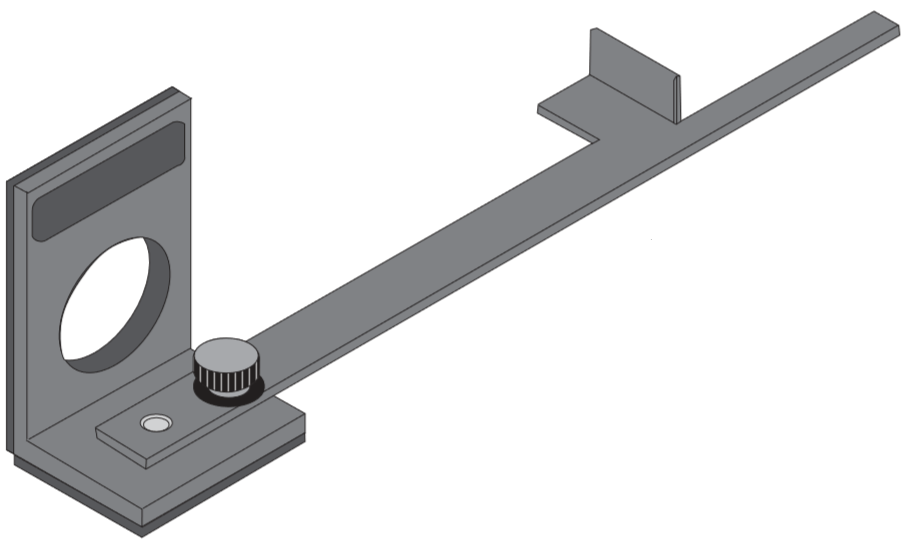}
  \caption{The Rotating Table consists of a component holder to place the glass plate on and an arm to manually rotate it.}
  \label{fig:6}
\end{figure}
\begin{figure}[!htbp]
  \centering
  \includegraphics[scale=0.5]{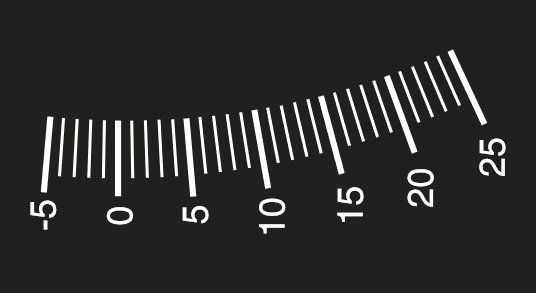}
  \caption{The vernier scale---which is positioned to the right of M$_2$ in Figure \ref{fig:1}--- is what we use to measure the angle which the arm has rotated the table. The alignment was set at $-0.40\degree\pm0.05$, so we add the positive value of that to all our angle measurements.}
  \label{fig:7}
\end{figure}

\begin{figure}[H]
\centering
\includegraphics[scale=0.5]{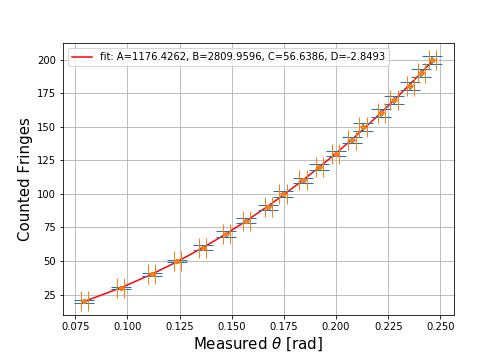}
\caption{Measured $N~\text{vs}~\theta$, where $0.4\degree$ have been added to $\theta$ as mentioned in Figure \ref{fig:6}. The angle has also been converted to radians. A curve $A\theta^3+B\theta^2+C\theta+D$ has been fitted to the data. The uncertainty in the parameters are $\delta A= 2097.06,~\delta B =1042.25,~\delta C=164.523,$ and $\delta D=8.169$.} 
\label{fig:8}
\end{figure}

\noindent While we rotated the arm we observe the viewing screen. Starting at 20, we read the Vernier scale every time 10 fringes pass our reference mark. Our measurements are plotted in Figure 
\ref{fig:8}.
\noindent The error for the measured angle is $0.05\degree$ as the Vernier scale is very small (after getting nearly identical results from a few trials we did manage to decrease the uncertainty). We also must take in consideration another $0.05\degree$ for the alignment offset uncertainty (as mentioned in Figure \ref{fig:6}), in total that sums up to $0.1\degree$. For $\delta N$ see Table \ref{tab:1}, as usual.

\section{Results}
\subsection{HeNe Laser Wavelength}
After gathering the measurements, we can calculate the uncertainty $\delta\lambda$ of \eqref{2} from the propagation of error formula \cite{Taylor}:
\begin{align}
\delta\lambda &= \sqrt{\left(\frac{\partial \lambda}{\partial N}\delta N\right)^2+\left(\frac{\partial \lambda}{\partial d_m}\delta d_m\right)^2} \nonumber \\
&=\sqrt{\frac{4{d_m}^2}{N^4}{\delta N}^2+\frac{4}{N^2}{\delta d_m}^2}\,.
    \label{5}
\end{align}
Since we have multiple measurements, we plot our data against the measured fringes $N$.
\begin{figure}[H]
\centering
\includegraphics[scale=0.5]{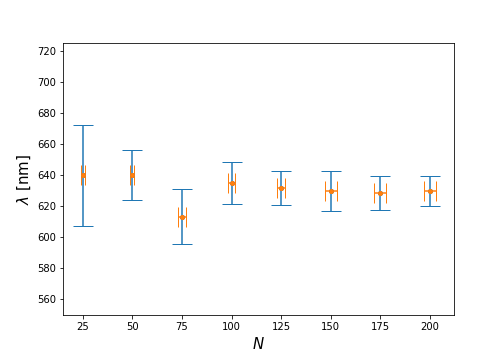}
\caption{Calculated $\lambda~\text{vs~measured}~N$. As the fringe count increases, $\lambda$ practically converges to $\lambda_0$ while its uncertainty decreases. Had a fit been included, it would have been a line with an almost zero slope and vertical axis intercept close to $\lambda_0$. Not included as it made each individual point hard to see.}
\label{fig:9}
\end{figure}

\noindent To get a final result from the wavelength of the laser beam, we take the weighted average $\overline\lambda$ of the individual wavelengths $\lambda_i$ in Figure \ref{fig:8};
\begin{align}
\overline{\lambda}=\dfrac{\sum \lambda_i w_i}{\sum w_i}\,,
    \label{6}
\end{align}
where the weight $w_i$ is,
\begin{align}
w_i=\frac{1}{\left(\delta\lambda_i\right)^2}\,.
    \label{7}
\end{align}
It follows that $\overline\lambda=630.6~\mathrm{nm}$. The predicted average error based on the weight is,
\begin{align}
\delta\overline\lambda=\frac{1}{\sqrt{\sum w_i}}\,,
    \label{8}
\end{align}
which comes out to 4.7 nm. However, the actual empirical spread of the calculated errors based off the measured data is:
\begin{align}
\sigma_\lambda=\sqrt{ \frac{1}{k} \sum_{i=1}^{k}\left(\lambda_i-\overline\lambda\right)^2}\,,
    \label{9}
\end{align}
where $k$ is the number of measurements \cite{FUN}. And the final wavelength is, $\overline\lambda\pm\sigma_\lambda$, so $\lambda=630.6\pm7.9~\mathrm{nm}$.

\subsection{Air's Index of Refraction}
\begin{figure}[H]
\centering
\includegraphics[scale=0.5]{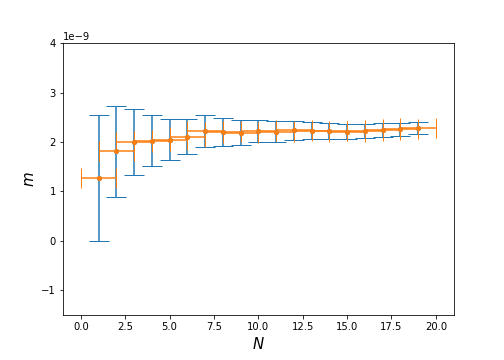}
\caption{Calculated $m$ vs measured $N$. With every measurement the error in the slope decreases. It also appears to converge to a value near 2.2 nPa$^{-1}$. To find each individual ${n_a}_i$, we must use each individual $m_i$ and $\delta m_i$ in \eqref{14}.}
\label{fig:10}
\end{figure}

From \eqref{2},
\begin{align}
N=\Delta N=2d\left(\frac{1}{\lambda_\alpha}-\frac{1}{\lambda_\beta}\right)\,,
    \label{10}
\end{align}
where the $\alpha$ and $\beta$ subscripts correspond to any two individual wavelengths. Then from \eqref{3}, 
\begin{align}
N=\frac{2d}{\lambda_0}\left(n_\alpha -n_\beta\right) \implies n_\alpha -n_\beta=\frac{N\lambda_0}{2d}\,.
    \label{11}
\end{align}
As mentioned in the introduction, $n$ varies linearly with $P$, so from the slope $m=\Delta n_a/\Delta P$ and \eqref{11},
\begin{align}
m=\frac{N\lambda_0}{2d\delta P}\,.
    \label{12}
\end{align}
Then, 
\begin{align}
n_a(P)=m\cdot P+n_0\,,
    \label{13}
\end{align}
where $n_0=1$ is the vacuum refractive index. It follows from this that
\begin{align}
    \delta n_a=\delta m\cdot P\,,
    \label{14}
\end{align}
where $\delta m$ is found from the propagation of uncertainty formula;
\begin{align}
\delta m &= \nonumber \\
& \frac{\lambda_0}{2d\delta P}\sqrt{\left(\delta n\right)^2 +\left(\dfrac{\delta d}{d}\right)^2 +\left(\dfrac{N\delta\Delta P}{\Delta P}\right)^2}\,.
\label{15}
\end{align}
We plot each individual $m_i$ and their uncertainties $\delta m_i$ in Figure \ref{fig:10}.

\vspace{-0.3cm}
\noindent Similar to the previous section, we have random uncertainties, so we need to do a statistical analysis. Replacing $\lambda$ with $m$ in equations \eqref{6} to \eqref{9}\footnote{The reason the equations have not explicitly been written out is to save space as well as clear redundancy. This is also the case for the next section.} we obtain the values in the following table.
\vspace{0.5cm}
\begin{table}[H] 
\centering
\caption{Obtained Averages for Slope [nPa$^{-1}$]}
\label{tab:2}
\begin{tabular}{c|c|c}
$\overline{m}$ & $\delta{m}$ & $\sigma_m$\\ \hline
\multicolumn{1}{c|}{2.228} & 0.046 & 0.256\\
\end{tabular}
\end{table}
\noindent Using the average slope $\overline{m}$ in Table \ref{tab:2} along with \eqref{13} we make a graph of the index of refraction of air against pressure. 
\begin{figure}[H]
\centering
\includegraphics[scale=0.5]{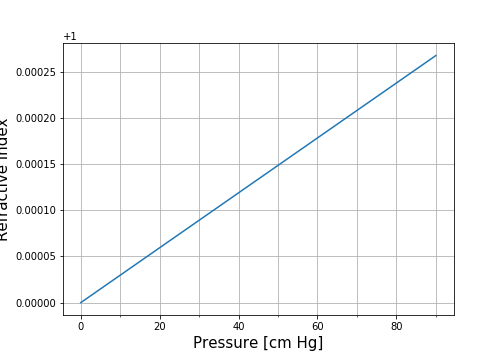}
\caption{Calculated $\overline{n}_a$ vs $P$. A range of 1000 pressures between 0 and 90 [cm Hg] was fed into \eqref{13} to create this graph.}
\label{fig:11}
\end{figure}
\noindent Extrapolating Figure \ref{fig:11} we find the refractive index of air to be 1.000226. From \eqref{14} and Table \ref{tab:2},
$\delta n_a$ is $5\times10^{-6}$. And, 
\begin{align}
    {\sigma_a}=\sigma_m\cdot P_{atm}\,,
    \label{16}
\end{align}
where $\sigma_a$ is the empirical spread for $n_a$ and $P_{atm}=101325$ Pa is the atmospheric pressure. So if, $$n_a=\overline{n}_a\pm\sigma_a\,,$$ then $n_a=1.000226 \pm 0.000026$.

\subsection{Glass's Index of Refraction}
Based on our measurements as well as the curve in Figure \ref{fig:8} we suspect that it will be difficult to find the change in path length. So we rely on the work done by Andrews, 1960 to analyze the content of \eqref{4} and write an equation for $n_g$ solely as a function of $\theta$ and $N$;
\begin{align}
n_g=\dfrac{\left(2t-N\lambda_0\right)(1-\cos\theta)+\left(\left(N\lambda_0\right)^2/\left(4t\right)\right)}{2t\left(1-\cos\theta\right)-N\lambda_0}\,,
    \label{17}
\end{align}
where $t$ is the thickness of the glass plate. For this we use the manufacturer value $t=6\pm0.5$ mm \cite{t}.\footnote{The exact part number for the glass plate is 003-04034. Since the manufacturer value is very precise the uncertainty is set to 5 in the next [unreported] significant figure.} 

\noindent Anyway, Since the second term in the numerator is negligible,
\begin{align}
n_g=\dfrac{\left(2t-N\lambda_0\right)(1-\cos\theta)}{2t\left(1-\cos\theta\right)-N\lambda_0}\,.
    \label{18}
\end{align}
Then from the propagation of error formula,
\begin{align}
\delta n_g = \hspace{7cm}\nonumber \\
 \frac{\lambda_0}{\zeta}\sqrt{\left(2t\eta\cos \theta\delta N\right)^2 +\left(2N\eta\cos\theta\delta t\right)^2 +\left(N\xi\sin
\theta\delta\theta\right)^2} ,
\label{19}
\end{align}
where $\xi$, $\zeta$, and $\eta$ are parameterized below to avoid clutterness in \eqref{19};
$$\begin{cases}
  \xi=N\lambda_0-2t\\
  \zeta=\left(\lambda _0 N+2t\cos \theta-2t\right)^2 \\
  \eta=\cos\theta-1 \, . 
\end{cases}$$
Plotting each ${n_g}_i$ with its error allows us to further analyze our data.
\begin{figure}[H]
\centering
\includegraphics[scale=0.5]{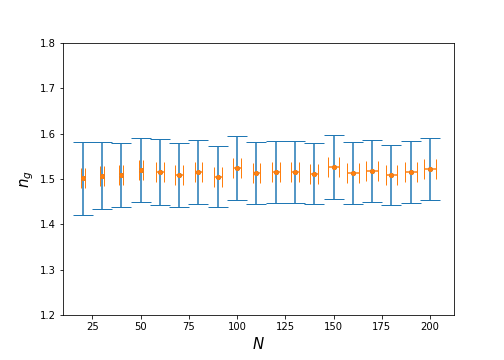}
\caption{Calculated $n_g$ vs measured $N$. Throughout the measurements the refractive index of glass appears to be consistently close to 1.5, no matter what the fringe count.}
\label{fig:12}
\end{figure}
\noindent Once again we must do a weighted analysis, so this time we replace $\lambda$ with $n_g$ in equations \eqref{6} to \eqref{9}.
\vspace{0.2605cm}
\begin{table}[H] 
\centering
\caption{Obtained Averages for $n_g$}
\label{tab:3}
\begin{tabular}{c|c|c}
$\overline{n}_g$ & $\delta{n_g}$ & $\sigma_g$\\ \hline
\multicolumn{1}{c|}{1.514} & 0.016 & 0.006\\
\end{tabular}
\end{table}
\noindent Our final index of refraction is $\overline{n}_g\pm\sigma_g$; so $n_g=1.514\pm0.006$.

\section{Discussion}
In the results of all three experiments we deduced both the calculated uncertainties of the weighted averages and the standard deviations of the measured errors relative to the weighted means. The former is an analytical prediction for the latter, and the latter is the verifiable error which we are mainly concerned with. Therefore, to evaluate the accuracy of our results, we use the following expression: 
\begin{align}
    \frac{\left|\Xi_{\mathrm{accepted}}-\Xi_{\mathrm{measured}}\right|}{\sigma} = \mathrm{std~away~from~\Xi_{\mathrm{accepted}}}.\footnotemark
\label{20}
\end{align}\par\footnotetext{As a reminder, $\sigma$ is the empirical spread, or as mentioned above, the 'verifiable error.'}

For the wavelength of the HeNe laser, our measured average result was $\lambda=630.6\pm7.9~\mathrm{nm}$. $0.28$ standard deviations away from $\lambda_0=632.8$ nm, agreeing with it. If we were to replace $\sigma_\lambda$ with $\delta\overline{\lambda}=4.7~\mathrm{nm}$, we would get $0.48$ standard deviations\footnote{If you were to plug in these \textit{exact} values in \eqref{20} you would get 0.4681 standard deviations. That is because our result have \textit{much} more significant figures than reported here. This applies for \textit{all} of our results ($\lambda,~n_a,~n_g)$.} away from the accepted value; also agreeing and implying that our errors were accurately estimated relative to our measurements.\par 
Our result of $n_a= 1.000226\pm0.000026$ is 1.44 standard deviations away from PASCO's measured 1.000263,\footnote{Since we are using PASCO's equipment, we will be comparing our result to their value as apposed to the accepted 1.000277 for 632.8 nm wavelengths.} hardly agreeing \cite{INTERFEROMETRY}. When in \eqref{20} we replace $\sigma_a$ with the predicted uncertainty $\delta\overline{n}_a=5e-6$, the standard deviation becomes 8.06, way off and not agreeing with PASCO's value. So even though our result with its original error agrees with the manufacturer, we highly underestimated our uncertainties.\par
What is more is that for the refractive index of crown glass we obtained $n_g=1.514\pm0.006$, 0.125 standard deviations away from the accepted value of 1.515\footnote{This value is explicitly for crown glass at 632.8 nm wavelengths.} \cite{Filmetrics}. If we instead use our calculated error based on our weighted average, $\delta\overline{n}_g=0.016$, our result would be 0.05 standard deviations away from the accepted value. Thoroughly agreeing with it.\par
Had we taken more measurements in the procedures for finding $\lambda$ and $n_g$ we may have obtained even smaller standard deviations, as the plotted points in Figures \ref{fig:9} and \ref{fig:10} would have converged closer to their respective accepted values. For the procedure of finding $n_a$: since in \eqref{15} $\Delta P$ and $d$ have been squared in the denominator relative to their respective uncertainties, the error that we underestimated was $\delta N$. Redoing the experiment and being more careful with our observation of the rings of Figure \ref{fig:2} we can get a better predicted uncertainty $\delta\overline{n}_a$ and hence a much better predicted standard deviation. That being said, all of our measurements and actual errors found from interferometery were successful and agreed with the accepted values.

{\small\singlespacing{

}}


\begin{thebibliography}{999} 

\bibitem{Wavelength}
Wavelength,
\emph{General Media},
\url{en.wikipedia.org/wiki/Wavelength},
Wikipedia.

\bibitem{nvsP}
  Kapoor \& Brown,
  \emph{Proceedings of the Third Symposium on Silicon Nitride and Silicon Dioxide Thin Insulating Films},
  The Electrochemical Society, Inc,
  1994.

\bibitem{3}
Monk,
\emph{Light Principles and Measurements},
McGraw-Hill,
1937.

\bibitem{INTERFEROMETRY}
PASCO,
\emph{Precision Interferometer}

\bibitem{Taylor}
  Taylor J.,
  \emph{An Introduction to Error Analysis, 2nd edition},
  University Science Books,
  1997.

\bibitem{FUN}
\emph{Standard Deviation and Variance},
\url{mathsisfun.com/data/standard-deviation.html},
Math is Fun.

\bibitem{Andrews}
Andrews, C.L.,
\emph{Optics of the Electromagnetic Spectrum},
Prentice-Hall,
1960. 

\bibitem{t}
PASCO,
\emph{Advanced Optics System},
\url{https://studylib.net/doc/8647418/advanced-optics-system},
page 13.

\bibitem{Filmetrics}
Filmetrics,
\emph{Refractive Index of BK7, Float Glass},
\url{filmetrics.com/refractive-index-database/BK7/Float-Glass}.

\end{thebibliography}
\end{document}